\documentclass[final,3p,times]{elsarticle}

\usepackage{amsfonts}
\usepackage{amsmath}
\usepackage{amssymb}
\usepackage{graphicx}
\usepackage{color}
\usepackage[normalem]{ulem}
\begin{document}
\begin{frontmatter}

\title{Spin glass induced by infinitesimal disorder in geometrically frustrated kagome lattice }

\author[ufsm]{M. Schmidt}\ead{mateusing85@gmail.com}
\author[ufsm]{F. M. Zimmer}\ead{fabiozimmer@gmail.com}
\author[iff]{S. G. Magalhaes}\ead{sgmagal@gmail.com}
\address[ufsm]{Departamento de Fisica, Universidade Federal de Santa Maria,
97105-900 Santa Maria, RS, Brazil}
\address[iff]{Instituto de Fisica, Universidade Federal Fluminense, 24210-346
Niter\'oi, RJ, Brazil}


\begin{abstract}
 We propose a method to study the magnetic properties of a disordered Ising kagome lattice. 
The model considers small spin clusters with infinite-range disordered couplings and short-range ferromagnetic (FE) or antiferromagnetic interactions. 
The correlated cluster mean-field theory is used to obtain an effective single-cluster problem. 
A finite disorder intensity in FE kagome lattice introduces a cluster spin-glass (CSG) phase. 
Nevertheless, an infinitesimal disorder stabilizes the CSG behavior in the geometrically frustrated kagome system. 
Entropy, magnetic susceptibility and spin-spin correlation are used to describe the interplay between disorder and geometric frustration (GF). 
We find that GF plays an important role in the low-disorder CSG phase.
However, the increase of disorder can rule out the effect of GF.
\end{abstract}

\end{frontmatter}

\section{Introduction}
Frustration in magnetism refers to competing exchange interactions that drive magnetic moments to an utterly conflicting situation \cite{Gardner2010,PRB_Moessner_2001}. There are two well known sources of frustration. 
The first one is caused by the geometry of some lattices combined with antiferromagnetic (AF) interactions in such a way that the ground state becomes highly degenerate without any conventional long-range order (LRO) \cite{lacroix2011introduction}. This can give rise to the spin-liquid state, in which the magnetic moments are extremely correlated but still fluctuate at lower temperatures \cite{Balents2010}.
The second one is caused by the presence of  disorder. For instance, it can produce competition between ferromagnetic (FE) and AF interactions arising a spin-glass (SG) state, where the magnetic moments are frozen in random directions \cite{Binder86}. Although geometric frustration (GF) is not required to stabilize the SG phase, there are several geometrically frustrated magnets showing SG-like behavior \cite{Gardner2010}. 
Intriguinly, in some of these materials the SG-like behavior appears at very low levels of disorder or even for what seems a disorder-free situation \cite{Gardner2010,
Booth2000,Ratcliff2002,LaForge2013}. These rather unexpected results raise issues such as how a SG-like state can be stabilized in such conditions. 

It should be remarked that there are few theoretical works that investigate how a SG-like state can be stabilised in geometrically frustrated systems at very low levels of disorder \cite{Saunders,Andreanov2010,Tam,Shinaoka,Gingras_2001}. In fact, one of the difficulties seems to be that the conceptual and mathematical framework to describe the disordered-driven SG state needs to  be adapted to describe this new class of problem. Recently, an important step forward has been made by Andreanov and collaborators \cite{Andreanov2010}. 
They formulated the problem in terms of a geometrically frustrated Edwards-Anderson (EA) type model. The disorder was introduced as a random deviation of the short-range uniform AF interactions. They showed that in this approach geometric frustration is quite sensitive to SG instability even for low disorder. 

Recently, a cluster spin-glass (CSG) phase \cite{Ratcliff2002, Chakrabarty_2014_2,Vijayanandhini_2009_PRB,Maji_2011} and short-range spin correlations within the SG-like phase \cite{Stewart_2011, Kindo2012, Clark_2014} have been found in several geometrically frustrated magnets. For example, the presence of spin clusters can favor a SG-like phase in Zn$_3$V$_3$O$_8$ \cite{Chakrabarty_2014_2}, CaBaFe$_{4-x}$Li$_x$O$_7$ \cite{Vijayanandhini_2009_PRB} and Nd$_5$Ge$_3$ \cite{ Maji_2011}. In particular, for ZnCr$_2$O$_4$, there is the suggestion that small spin clusters rather than canonical spins can bring hypersensitivity to disorder \cite{Ratcliff2002}. 
In addition, the kagome Ising antiferromagnet Co$_3$Mg(OH)$_6$Cl$_2$ provides a SG-like behavior with short-range correlations, apparently without disorder \cite{Kindo2012}.  In this magnet the neutron diffraction and muon spin rotation/relaxation results could also support the presence of small spin clusters.  In these systems the freezing 
behavior may not be well described as a conventional SG. From the theoretical side, novel techniques are then needed to account for the cluster SG state found in geometrically frustrated systems.

More recently, new approaches
have been proposed using clusters of spins instead of canonical ones
\cite{Yokota2014,prezimmer2014,schmidt_script2015}.
Some of these theories (see, for instance, Refs. \cite{prezimmer2014,schmidt_script2015}) follow the procedure introduced by Soukoulis and Levin \cite{Soukoulis78-1,Soukoulis78-2}.
The Soukoulis-Levin theory has been proposed as an improvement of the mean-field description of the SG problem by introducing intracluster degrees of freedom. 
Their Hamiltonian has two terms, the first one gives the intracluster short-range spin interactions which can be AF or FE. In this term, details of the original lattice can be included in the cluster geometry. The disorder is given by a long-range random intercluster (in fact, an infinite range) interaction from which a cluster SG state can appear. 
However, the procedure to describe the AF/FE LRO in the Soukoulis-Levin theory is rather artificial. The AF/FE LRO is obtained by displacing the distribution of the long-range random intercluster interactions by a uniform term, which is completely unrelated to intracluster AF/FE short-range interactions. For a geometrically frustrated lattice such procedure implies that GF effects coming from intracluster spin correlations are poorly connected with intercluster spin correlations. In view of that, it is crucial to account better for the role of AF interactions in this particular type of disordered lattice. 

The purpose of the present work is to study how a CSG phase can be stabilised in the disordered kagome lattice Ising model for low levels of disorder. The kagome lattice is one of the most straightforward examples of highly geometrically frustrated system being a possible realization of exotic magnetic states \cite{lacroix2011introduction,Han2012, Yan03062011}. We follow a procedure similar to that of the Soukoulis-Levin theory. The lattice is divided in a set of clusters with short-range AF/FE intraclusters interactions. Nevertheless, the intercluster term does not have only the long-range random interactions but also short-range AF/FE interactions. Here, we adopt the correlated cluster mean-field (CCMF) theory to treat these short-range interactions \cite{yamamoto}. In particular, the Soukoulis-Levin theory combined with the CCMF method have been used in the disordered square lattice Ising model showing that a LRO AF evolves directly from the AF short-range interactions below the N\'eel temperature (see Ref. \cite{ZimmerSchmidt2014}). 
In addition, at very low temperature, these short-range AF interactions produce a spin cluster compensation which makes less effective the random intercluster interaction to stabilize a CSG phase \cite{ZimmerSchmidt2014}. On the contrary, for the disordered kagome lattice,  it is expected that GF effects coming from the cluster core should undermine the AF LRO even at very low temperature. However, they also could enhance the uncompensation of the spin cluster, which makes the random intercluster interaction more effective. As a consequence, the level of disorder which stabilizes the CSG would be lowered. In order to check how effective this mechanism is, we also analyse the problem when the AF interactions are replaced by FE ones. In this case, a FE LRO can appear, which should change the level of disorder in which the CSG phase is stabilized as compared with the AF case. 

The CCMF theory has been initially proposed as a procedure to improve the mean-field approximation for canonical spins taking into account spin correlations neglected by other mean field approaches \cite{yamamoto}. In the CCMF theory the lattice is divided into symmetric clusters which are, in fact, just an artefact to incorporate spins correlations. The internal field acting on a given cluster is strongly dependent on the surface spin configurations of this cluster. The CCMF theory is able to improve results for critical temperature, spin-spin correlations and critical exponents in FE spin systems as compared, for instance, with usual mean-field approximations \cite{yamamoto, Yamamoto2}. It is important to remark that in the present work we have verified whether the CCMF theory applied in the clean AF kagome lattice Ising model can correctly capture GF effects. The results are in good agrement with the exact ones  (see Ref. \cite{Syozi1951}), such as absence of LRO  at finite temperatures and residual entropy for the geometrically frutrated case.

We also adopt the infinite-range intercluster random interaction given as the van Hemmen (vH) model \cite{vanHemmen82,vanHemmen83}.
The original vH model for canonical spins belongs to a class of separable on-site SG model. It is a weakly correlated disorder that introduces a small amount of frustrated configurations and, most importantly,  it can be treated avoiding the replica method. This model, even lacking the multiplicity of metastable states found in the SK model,   
gives  several static quantities in good agreement with real SG \cite{vanHemmen82,vanHemmen83}.
In particular, it allows accessing thermodynamic quantities, such as entropy, inside the SG phase without trouble caused by the replicas. It should be mentioned that the vH disordered cluster model in a square lattice treated with the CCMF method has been studied in Ref. \cite{ZimmerSchmidt2014}.  This allows to compare two distinct lattice geometries, one of them with GF, both solved with the same method.

The remainder of the paper is organized as follows. The disordered kagome lattice in a CCMF approach is introduced in Sec. \ref{model}. Numerical results for the kagome lattice without disorder are presented in Sec. \ref{CCMF}. The interplay between disorder and GF is discussed in Sec. \ref{vH}. Remarks and conclusion are given in Sec. \ref{conc}.

\section{Model}\label{model}
\begin{figure*}
\center{\includegraphics[angle=0,width=0.95\textwidth]{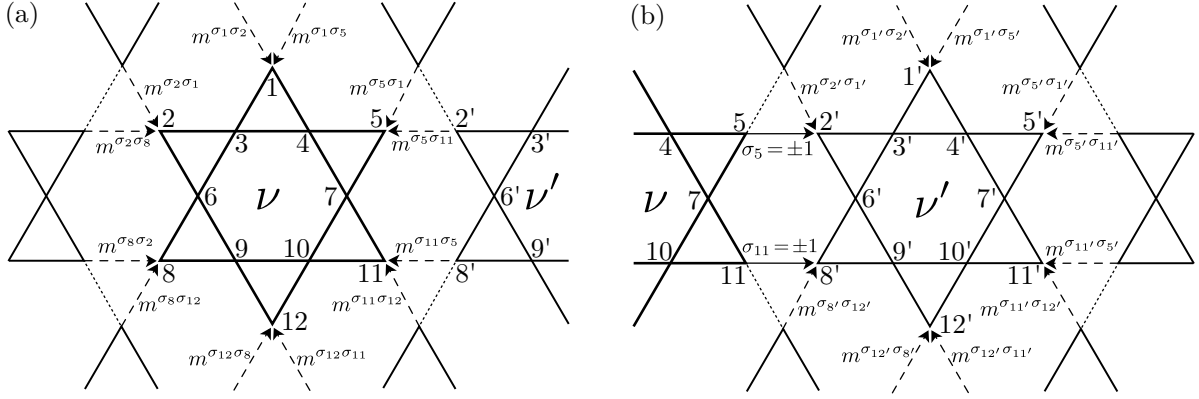}}
\caption{Schematic representation for a kagome lattice divided into clusters with $n_s=12$. 
The  mean fields are pointed by arrows that represent (a) the interactions between cluster $\nu$ and its neighbors; 
(b) the interactions on $\nu'$ used to evaluate the $m^{s\bar{s}}$. }
\label{square_4}  \end{figure*}

 We start from the Ising model $H=-\sum_{i,j} J_{ij}\sigma_i\sigma_j $
on the kagome lattice, where $J_{ij}=J_{0}+\delta J_{ij}$ is the exchange interaction with a random deviation $\delta J_{ij}$. The lattice is divided into $N_{cl}$ star-shaped clusters with $n_s=12$ sites (see Fig. (\ref{square_4})), resulting in
\begin{equation}
H=-\sum_{\nu}^{N_{cl}}\sum_{i,j}^{n_{s}} (J_{0}^{\nu\nu}+\delta J_{ij}^{\nu\nu})\sigma_{\nu_{i}}\sigma_{\nu_{j}}-\sum_{\nu,\lambda}^{N_{cl}}\sum_{i,j}(J_{0}^{\nu\lambda}+\delta J_{ij}^{\nu\lambda})\sigma_{\nu_{i}}\sigma_{\lambda_{j}}
\label{hamiltonian01}
\end{equation}
where $\sigma_{\nu_i}=\pm 1$ is the Ising spin of site $i$ of cluster $\nu$.
The first term of Eq. (\ref{hamiltonian01}) represents intracluster interactions, in which we consider only FE or AF short-range interactions, $J_{0}^{\nu\nu}=J_0$, without random deviation ($\delta J^{\nu\nu}_{ij}=0$). For the  second one, we assume two types of couplings: the $J_0$ between nearest-neighbor spins of neighbor clusters and long-range disordered couplings among all spin pairs of different clusters ($\delta J^{\nu\lambda}_{ij}=\delta J^{\nu\lambda}$).
Explicitly, the Hamiltonian is represented by:
\begin{equation}
H=-\sum_{\nu}^{N_{cl}}\sum_{(i,j)}^{n_{s}}J_0\sigma_{\nu_{i}}\sigma_{\nu_{j}}-\sum_{(\nu_i,\lambda_j)}J_0\sigma_{\nu_{i}}\sigma_{\lambda_{j}} 
-\sum_{\nu,\lambda}^{N_{cl}}\delta J^{\nu\lambda}S_{\nu}S_{\lambda}, \label{hamiltonian}
\end{equation}
where $(\nu_i,\lambda_j)$ denotes sum over nearest-neighbor spins of different clusters, and $S_\nu=\sum_{i}^{n_s}\sigma_{\nu_i}$ represents the total magnetic moment of cluster $\nu$.  In other way, model (\ref{hamiltonian}) is a kind of hybrid between a cluster approach (CCMF) to the kagome lattice, which allows to take into account short-range interactions, and the introduction of disordered couplings between these clusters, which could induce a SG behavior. 
In this cluster formalism the lattice structure is preserved inside the clusters, in which the short-range interactions remain relevant to determine the system behavior, mainly in the geometrically frustrated case.
 In particular,
the intercluster disorder follows the van Hemmen coupling \cite{vanHemmen83}: $\delta J^{\nu\lambda}=\frac{J}{N_{cl}n_s}(\xi_{\nu}\eta_{\lambda}+\xi_{\lambda}\eta_{\nu}),$ where $J$ is the strength of disorder, and $\xi_{\nu}$'s and $\eta_{\lambda}$'s are independent random variables with Gaussian distributions of variance one.

The intercluster disorder is evaluated as in Ref. \cite{ZimmerSchmidt2014}. The last term of Eq. (\ref{hamiltonian}) is written in quadratic forms. The partition function for a set of quenched variables $\{\xi,\eta\}$ can then be obtained as 
\begin{equation}\begin{split}
Z(\xi,\eta)= \mbox{Tr} ~e^{-\beta H(\xi,\eta)}=\int  dq_1 dq_2 dq_3
\exp \left\{
-N\left[\frac{\beta J(q_3^2-q_1^2-q_2^2)}{2}-\frac{1}{N} \ln \mbox{Tr} \exp(-\beta H_{eff})\right]\right\} \end{split}\label{h_sg}\end{equation}
where $q_1(\xi,\eta)$, $q_2(\xi,\eta)$, and $q_3(\xi,\eta)$ are 
introduced by Hubbard-Stratonovich transformations,  $N=N_{cl}n_{s}$ and 
$H_{eff}=-J\sum_{\nu}^{N_{cl}}([(\xi_\nu+\eta_\nu)q_3-\xi_\nu q_1-\eta_\nu q_2] S_\nu-J_{0}\sum_{(i,j)\in\nu}\sigma_{{i}}\sigma_{j}) -\sum_{(\nu_i,\lambda_j)} {J}_{0}\sigma_{\nu_{i}}\sigma_{\nu_{j}}. $ 
In the thermodynamic limit ($N_{cl}\rightarrow\infty$), the functional integrals over $\{q_n(\xi,\eta)\}$ ($n = 1, 2$ and 3)  are computed by the steepest descent method.  As results $q_3=q_1+q_2$ and $q=q_1=q_2$, in which the SG order parameter $q$ is obtained from \cite{vanHemmen83,vanHemmen82}
\begin{equation} \begin{split}
q=\frac{1}{N}\left\langle \frac{\mbox{Tr} \sum_{\nu} \frac{(\xi_\nu+\eta_\nu)}{2} S_\nu e^{-\beta H_{eff}}}{\mbox{Tr} ~e^{-\beta H_{eff}}}\right\rangle_{\xi,\eta}, 
\label{sgOP}\end{split}\end{equation}
where 
\begin{equation}\begin{split}
H_{eff}=\sum_{\nu}^{N_{cl}}[-J(\xi_\nu+\eta_\nu)q S_\nu-\sum_{(i,j)\in\nu}J_{0}\sigma_{{i}}\sigma_{j}] 
-\sum_{(\nu_i,\lambda_j)} {J}_{0}\sigma_{\nu_{i}}\sigma_{\nu_{j}}
\label{1hamiltonian_eff}
\end{split}\end{equation}
and $\langle\cdots\rangle_{\xi,\eta}$ represents the average over $\xi$ and $\eta$. 
In this approach, the disorder determines the cluster-like behavior, in which the cluster magnetic moment is under an effective  internal field that depends on the SG order parameter $q$ (see first term of Eq. (\ref{1hamiltonian_eff})). In particular, $q$ is an order parameter that represents the behavior of all cluster spins, which characterizes a CSG phase.    

In Eq.  (\ref{1hamiltonian_eff}) there are still short-range couplings (FE or AF) between neighbor clusters. 
These spin interactions of different clusters are treated in the framework of the CCMF approach \cite{yamamoto},
which results in the following effective one-cluster model
\begin{equation}\begin{split}
H_{eff}= -\sum_{(i,j)\in \nu} J_{0}\sigma_{i}\sigma_{j}-\sum_{i\in\nu} J(\xi_\nu+\eta_\nu)q \sigma_{i}
 -\sum_{i\in \bar{\nu}}h^{eff}_{i}\sigma_{i}
\label{hamiltonian_eff},
\end{split}\end{equation} 
where $h^{eff}_{i}=J_0(m^{\sigma_i \sigma_j}+m^{\sigma_i \sigma_k})$ is an effective field acting on the spin $\sigma_{i}$ of site $i$, which belongs to the cluster border ($\bar{\nu}$).  
The indices $j$ and $k$ represent sites of $\bar{\nu}$ and nearest-neighbors of site $i$.
Hereafter, we remove the spin cluster index $\nu$. In the CCMF routine the mean fields $m^{\sigma_i \sigma_j}$ and $m^{\sigma_i \sigma_k}$ are dependent of the spin configurations of cluster $\nu$. For each spin configuration, $\{\sigma_i \sigma_{j}\} = \{++,+-,-+,--\}$, a different mean field is obtained: $m^{++}$, $m^{+-}$, $m^{-+}$, $m^{--}$. 
For example, the set of mean fields $\{m^{\sigma_{5} \sigma_{11}}\}$ acting on spin $\sigma_5$ is obtained from the average value of $\sigma_{2'}$ of cluster $\nu'$ for all possible configurations of the spins $\sigma_{5}$ and $\sigma_{11}$  of cluster $\nu$ (see Fig. \ref{square_4}). This average is evaluated from
\begin{equation}
 m^{s\bar{s}}  
=\left\langle \frac{{\rm Tr} \sigma_{2'} e^{-\beta H'_{eff}} }{{\rm Tr} e^{-\beta H'_{eff}}}\right\rangle_{\xi,\eta},
\label{mean_field}\end{equation}
where $s$ and $\bar{s}$ represent the spin states $\sigma_5$ and $\sigma_{11}$, respectively. The Hamiltonian of cluster $\nu'$ is given by
\begin{equation}\begin{split}
H'_{eff}= -\sum_{(i,j)\in \nu'} J_0 \sigma_{i}\sigma_{j}
-\sum_{i \in \nu'} J(\xi+\eta)q \sigma_{{i}}
-\sum_{\begin{subarray}{c}i \in \nu'\\\left(i\neq 2',8'\right) \end{subarray}}h^{eff}_{i}\sigma_{{i}}
- J_0\sigma_{2'}(s+m^{\sigma_{2'} \sigma_{1'}}) 
- J_0\sigma_{8'}(\bar{s}+m^{\sigma_{8'} \sigma_{12'}}),
\label{hamiltonian_eff2}
\end{split}\end{equation} 
where the mean fields $m^{\sigma_{2'} \sigma_{8'}}$ and $m^{\sigma_{8'} \sigma_{2'}}$ are replaced by $s$ and $\bar{s}$, respectively.
Here we can explore the symmetry of the spin cluster border to obtain all sets of mean fields.
We checked the validity of this symmetry by explicitly evaluating all mean fields for all border spins (without any symmetry). The result was the same as that  obtained by adopting the symmetry.

The effective one-cluster problem is solved  by computing Eqs. (\ref{sgOP}), (\ref{mean_field}) and (\ref{hamiltonian_eff2}) self-consistently. 
After, we can derive other thermodynamic quantities. For instance, the magnetization per site and the magnetic susceptibility $\chi$ are calculated as 
\begin{equation}
 m =\frac{1}{n_s}\left\langle \frac{{\rm Tr} \sum_{i}^{n_s}\sigma_{i} e^{-\beta H_{eff}} }{{\rm Tr} e^{-\beta H_{eff}}}\right\rangle_{\xi,\eta}
\label{magn}\end{equation}
and $\chi=\left(\frac{\partial m}{\partial h}\right)_{h=0}$. We also compute the staggered magnetization of three sublattice to identify a possible AF order.The internal energy per site is obtained from 
\begin{equation}\begin{split}u=-\frac{J_0}{n_s}  \left\langle  \frac{ {\rm Tr} \sum_{(i,j) \in \nu } \sigma_i \sigma_j e^{-\beta H_{eff}}}{{\rm Tr} e^{-\beta H_{eff}}} \right\rangle_{\xi,\eta}-
 \frac{1}{2n_s}\left\langle  \frac{ {\rm Tr} \sum_{i \in \bar{\nu} } h_{i}^{eff}\sigma_i e^{-\beta H_{eff}}}{{\rm Tr} e^{-\beta H_{eff}}} \right\rangle_{\xi,\eta}-J q^2,
\label{internal_ene}\end{split}\end{equation}
where the factor $1/2$ accounts for the fact that each effective interaction between clusters is shared. When $J$ or $q$ is equal to zero, the cluster is an artefact to introduce correlations and the better expression for the internal energy is $u=-\frac{4}{3} \frac{J_0}{n_s} \left\langle \frac{ {\rm Tr} \sum_{(i,j) \in \nu } \sigma_i \sigma_j e^{-\beta H_{eff}}}{{\rm Tr} e^{-\beta H_{eff}}} \right\rangle_{\xi,\eta} $, where the factor $4/3$ accounts for the intercluster interactions  \cite{Wysin2000}. 
In particular, the first term between brackets of Eq. (\ref{internal_ene})  corresponds to the intracluster spin-spin correlation function $\langle \sigma_i\sigma_j \rangle_{intra}$. The specific heat $(c_v)$ is obtained by numerical derivation of the internal energy. The entropy per site is given by integration of $c_v/T$: 
\begin{equation}
{s}{}=\int_{0}^{T} \frac{c_v}{T'}dT'=\ln(2)-\int_{T}^{\infty} \frac{c_v}{T'}dT'.
\end{equation}

\begin{figure}
\center{\includegraphics[angle=0,width=0.49\textwidth]{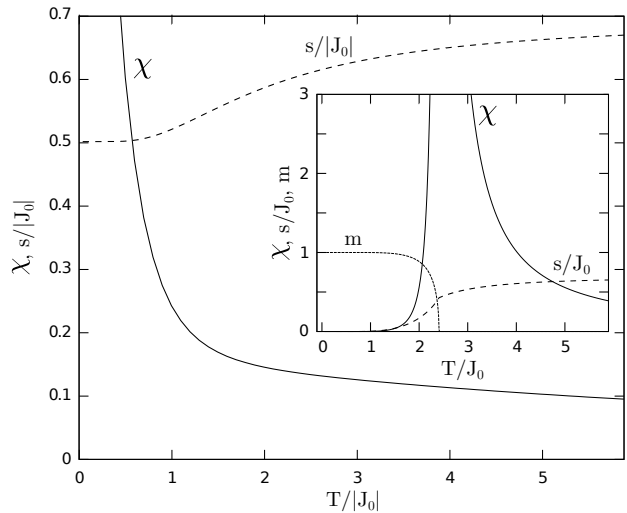}}
\caption{Entropy and magnetic susceptibility as a function of the temperature for the geometrically frustrated kagome lattice without disorder, in which $s(T\rightarrow 0)=0.503$. 
The inset exhibits the behavior of magnetization, $\chi$, and $s$ 
for ferromagnetic interactions.  }
\label{kagome_pure}  \end{figure}

\section{Results}
Numerical results are obtained from the exact diagonalization of the effective single-cluster problem.  We obtain other thermodynamic quantities such as magnetization, magnetic susceptibility, entropy, and spin-spin correlation function. We also build phase diagrams of the temperature $T$ versus disorder intensity $J$.  The results are in unity of $|J_0|$, which for numerical purposes is used $|J_0|=1$. 
The cluster spin-glass (CSG) phase occurs when $q\neq0$ and $m = 0$, and the ferromagnetic (FE) order is characterized by $m\neq0$ and $q=0$. The antiferromagnetic order is not found.

\subsection{Kagome lattice within the CCMF}\label{CCMF}
Figure \ref{kagome_pure} presents the results for the pure kagome lattice ($J=0$). It exhibits the entropy $s$ and magnetic susceptibility $\chi$ for the geometrically frustrated case ($J_0<0$). The entropy curve goes towards a non-zero value as $T \rightarrow 0$: $S(T\rightarrow 0)=0.503$. 
This residual entropy is a consequence of the strong GF and indicates a high ground-state degeneracy. 
This finding is in perfect agreement with the exact result for the AF Ising kagome lattice \cite{Kano1953}. 
Furthermore, the magnetic susceptibility diverges as $T \rightarrow 0$; however, there is no phase transition at finite temperature \cite{Syozi1951}. 
The residual entropy and the absence of LRO suggest a spin-liquid ground state \cite{Balents2010} that is in agreement with early results \cite{Loh_PRB}.
Therefore, the CCMF theory arises as a good alternative to study this  strongly  geometrically frustrated system.
In addition, the inset of Fig. \ref{kagome_pure} shows results for thjae FE kagome lattice ($J_0>0$). The magnetization and susceptibility indicate a paramagnetic PM-FE phase transition at $T_c=2.41 J_0$ that overestimates around $ 12\%$ the exact solution ($T_{c}^{exact} = 2.14 J_0$ \cite{Kano1953}). 
The entropy also exhibits a typical curve for the Ising FE kagome lattice. It is important to remark that the CCMF theory applied to the kagome lattice provides better results (as compared with the exact solution) than conventional mean-field theories \cite{Takeo}.

\subsection{Kagome lattice with disordered clusters}\label{vH}
\begin{figure*}
\center{\includegraphics[angle=0,width=0.49\textwidth]{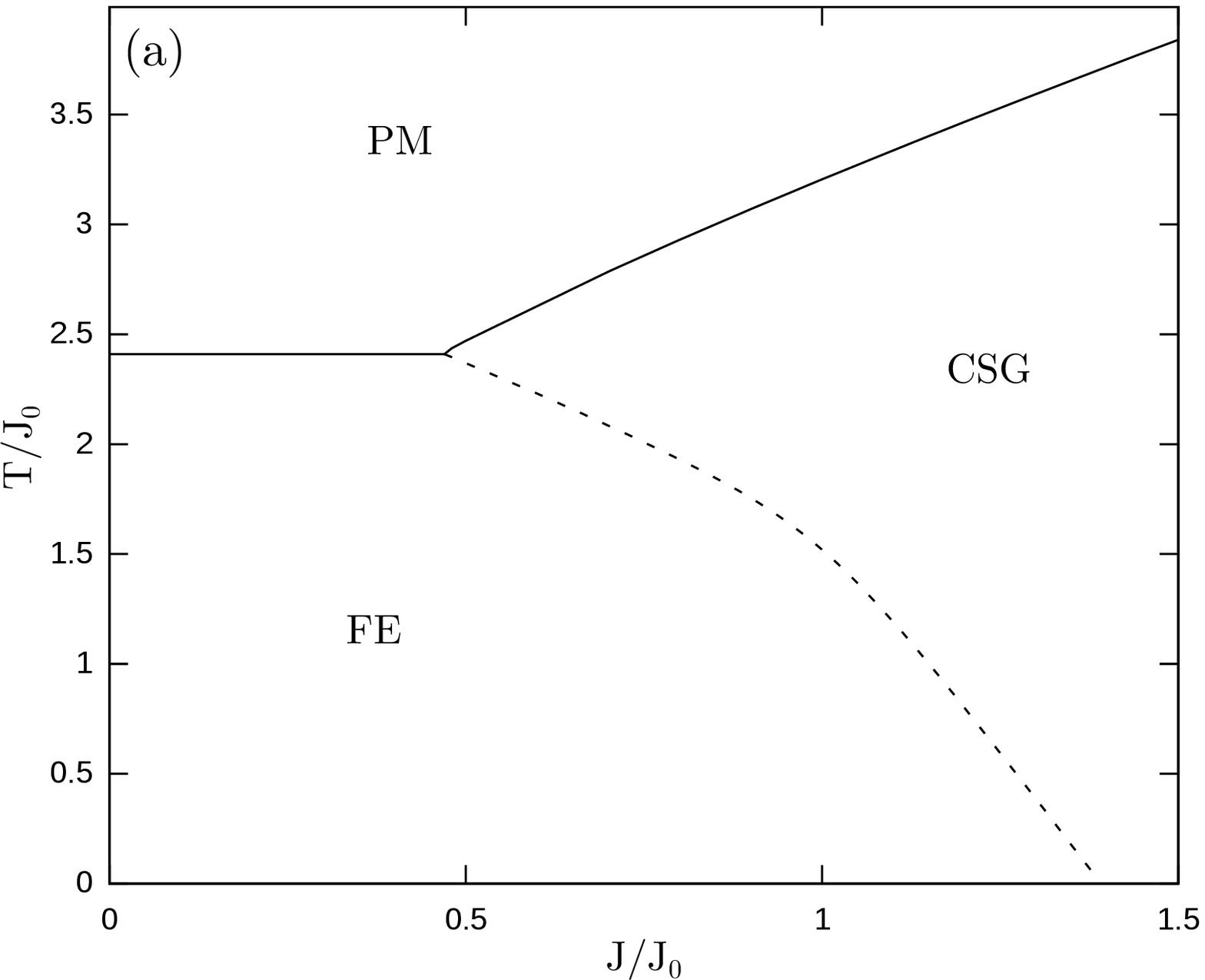}\includegraphics[angle=0,width=0.49\textwidth]{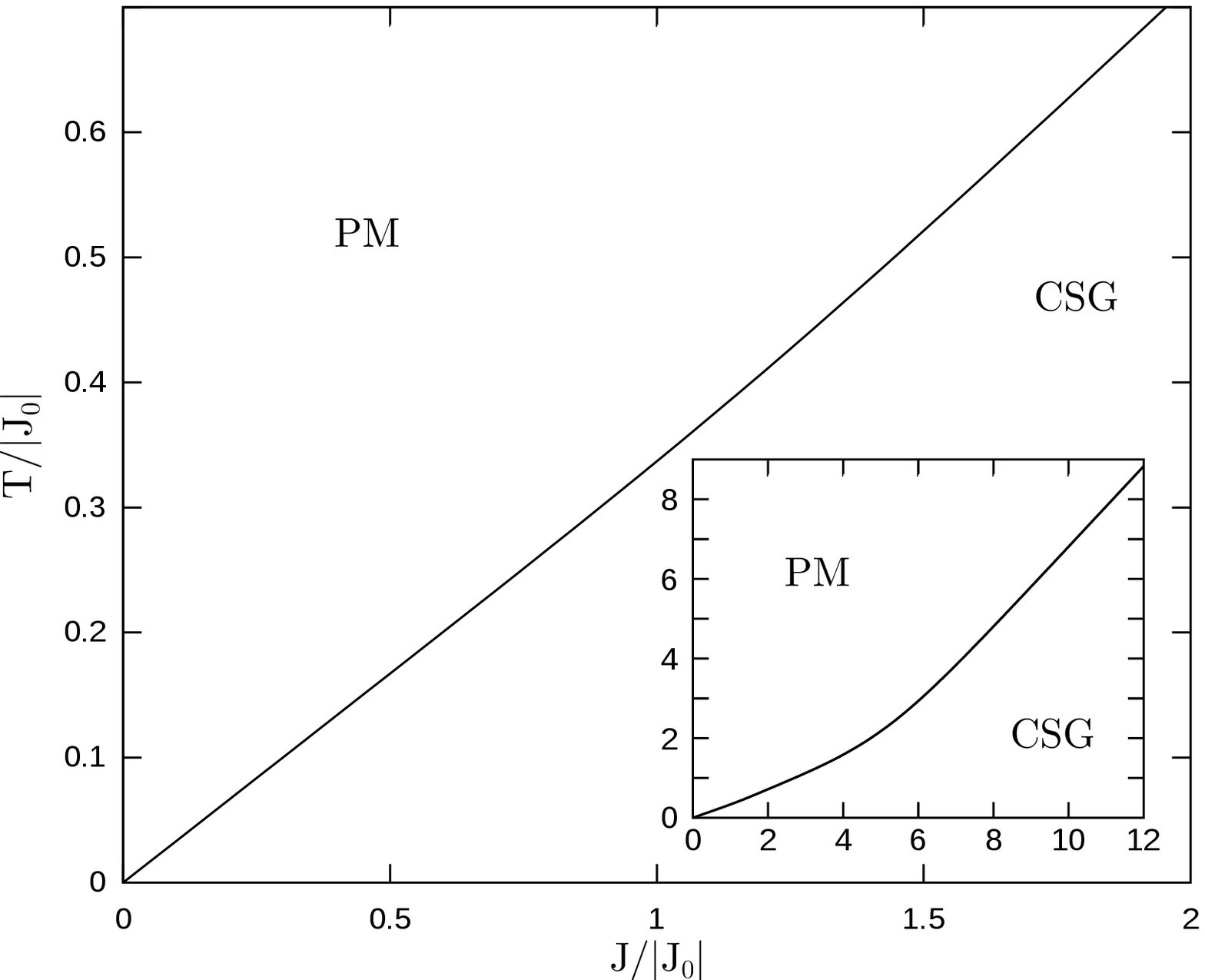}}
\caption{Phase diagrams of the critical temperatures as a function of the disorder intensity. (a) Ferromagnetic interactions. (b) Antiferromagnetic interactions. The inset exhibits the $T_f$ behavior for high disorder intensities.}
\label{phase_diag}  \end{figure*}

The effect of disorder on the magnetic properties of the kagome lattice is now analyzed. For instance,  the FE long-range order is found  in the case  without GF ($J_0>0$). Figure (\ref{phase_diag}-a) shows that the FE order is still stable for low intensities of disorder. However, the CSG behavior appears when $J$ is higher than a critical value. There is a range of disorder in which a reentrant PM-CSG-FE phase transition occurs when the temperature decreases. In particular, the CSG-FE transition is first-order and it is located by comparing free energies obtained from: $F=U-TS$. To summarize the CSG phase occurs only at high disordered regimes in the kagome lattice without GF. 

Figure (\ref{phase_diag}-b) exhibits $T_f$ as a function of $J$ for the AF kagome lattice, in which the interplay between GF and disorder  brings on a completely different scenario.  There is no LRO at finite temperature for $J = 0$, but an infinitesimal disorder is enough to change the spin-liquid ground state into the CSG phase. 
It means that the GF helps introducing the CSG phase at very small intensities of disorder. Furthermore, there are clearly two distinct behaviors for $T_f(J)$. One for weak disorder, in which the effect of GF is relevant. Other for high intensities of disorder, where the $T_f(J)$ inclination  is more pronounced (see inset of Fig. \ref{phase_diag}-b).

\begin{figure}[htb]
\center{\includegraphics[width=0.49\textwidth]{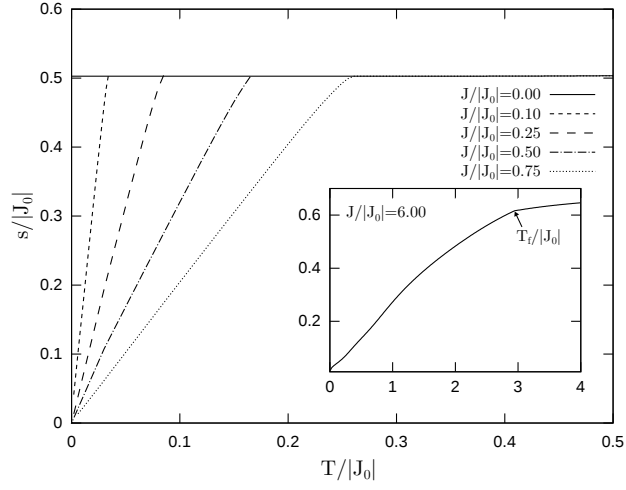}}
\caption{Entropy as a function of the temperature for  the increase of disorder intensity in the AF case. The inset shows a result for a high value of $J$.   }
\label{entropy}  \end{figure}

In the following, the interplay between GF and disorder is discussed in terms of some thermodynamic quantities. For instance, the entropic behavior of the AF kagome lattice is strongly affected by disorder. Figure \ref{entropy} shows that a very low disorder intensity leads the entropy to zero when $T \rightarrow 0$. This indicates that an infinitesimal disorder is able to vanish the residual entropy and stabilize the CSG phase. It is important to point that the CSG phase is between clusters. In fact, the system ground state becomes cluster CSG. In other words, the entropy is an extensive quantity that characterizes the whole system, but it can not identify spin states inside the clusters.  For higher disorder intensities, the entropy also decreases towards zero when the temperature diminishes (see the inset plot of Fig. \ref{entropy}).

\begin{figure*}[htb]
\center{\includegraphics[width=0.48\textwidth]{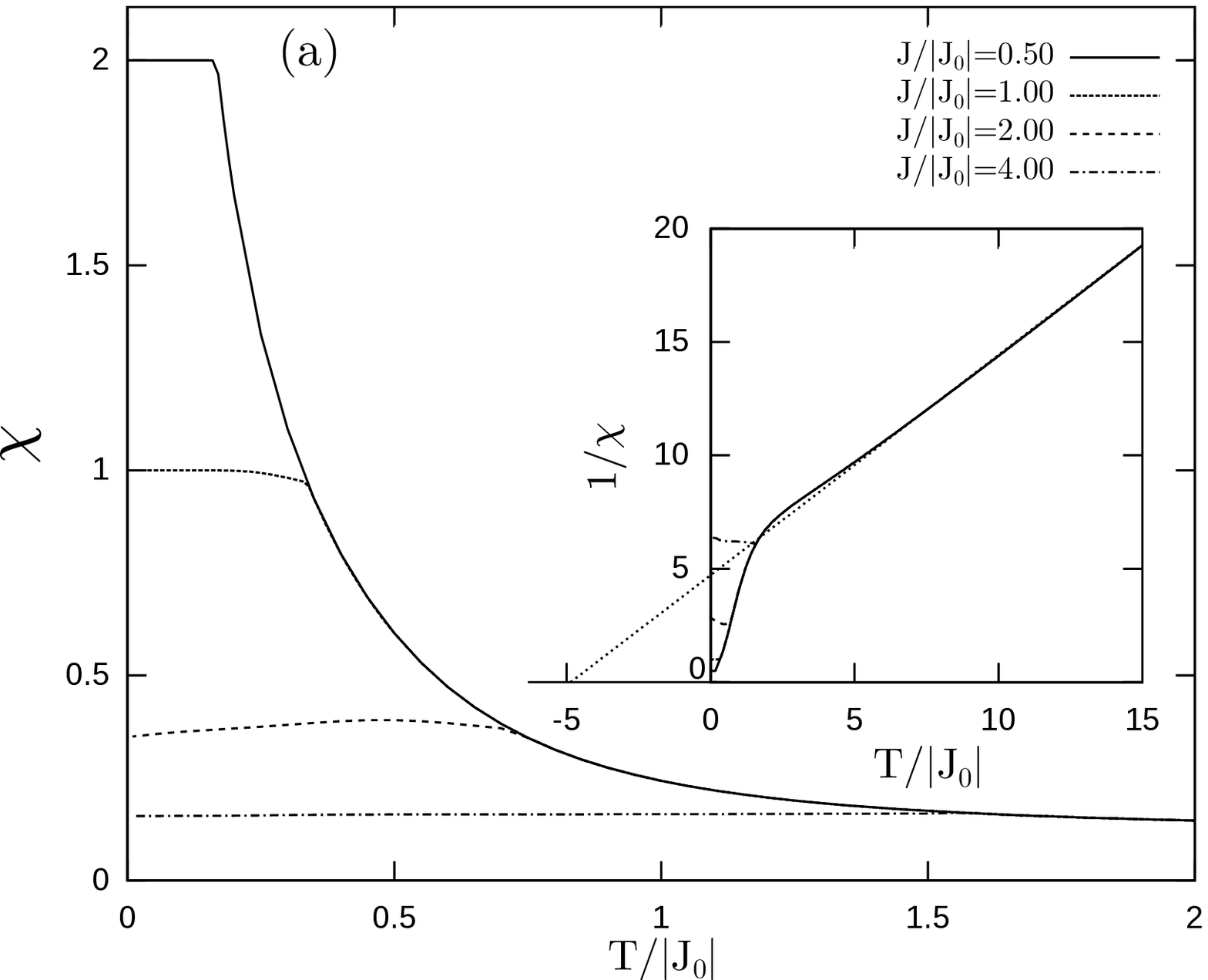}
\includegraphics[width=0.48\textwidth]{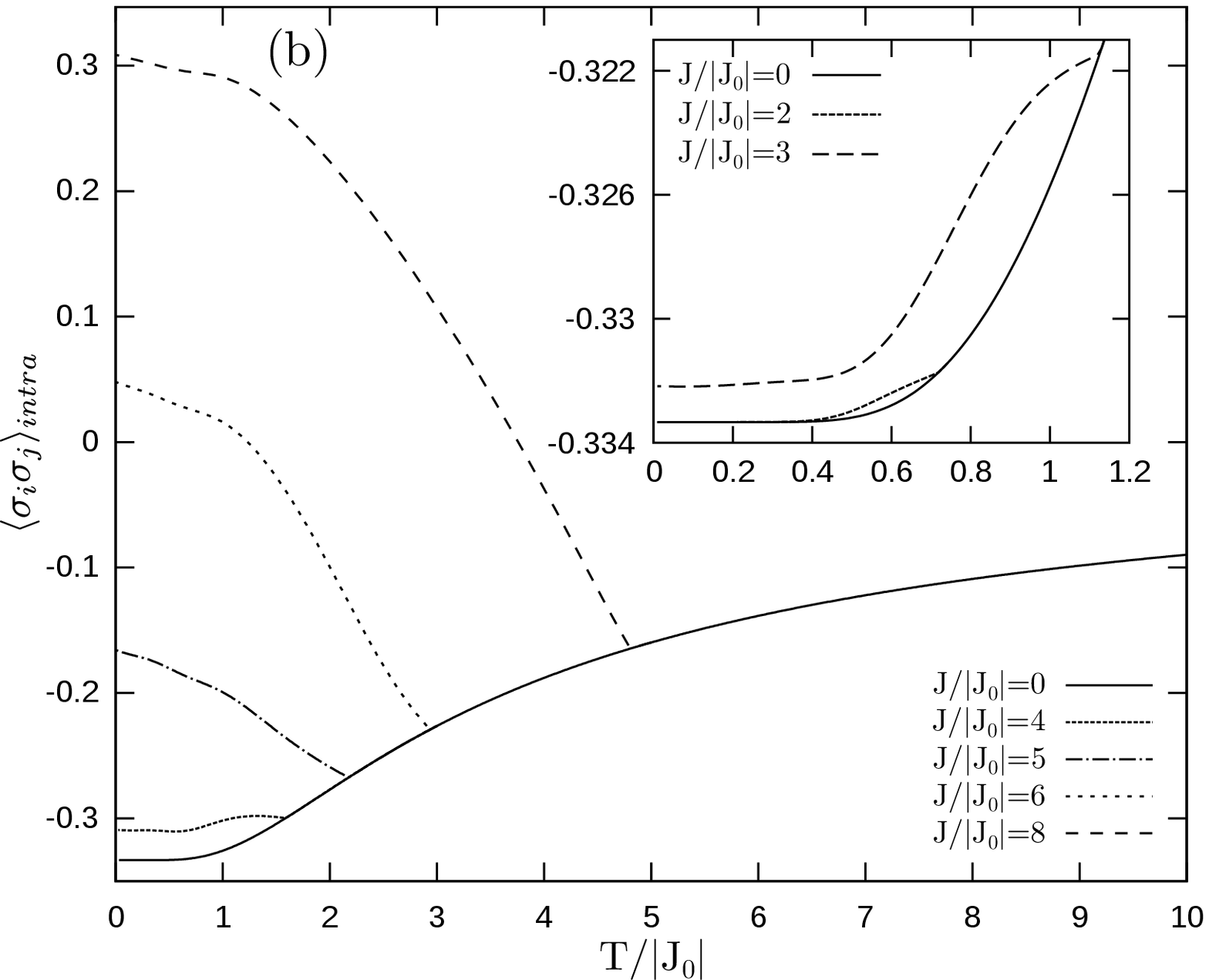}}
\caption{(a) Magnetic susceptibility versus temperature for several values of $J$. The inset exhibits the inverse of susceptibility and the dotted line stands for the linear extrapolation from high temperatures. (b) Spin-spin correlation function behavior for different intensities of disorder. The inset shows results for lower $J$ values.    }
\label{corr}  \end{figure*}

Figure \ref{corr}-a shows the magnetic susceptibility, which presents a flat plateau below $T_f$ that is a characteristic of  spin glass phase.  For high temperatures, the system follows the Curie-Weiss law. These features are found for any finite disorder. The linear extrapolation of $\chi^{-1}$ (see the inset in Fig. \ref{corr}-a) allows to estimate the Curie-Weiss temperature $\theta_{CW} \approx -5 |J_0|$ that is independent of $J$. In the present approach, the  $\chi$ of the PM phase is not affected by the disorder intensity (see Fig. \ref{corr}-a). As a consequence $\theta_{CW}$ becomes independent of $J$. However, an important result is obtained from the frustration parameter $f_p=|\theta_{CW}|/T_f$ when the disorder changes. For a weak disorder regime, the frustration parameter is higher than for the strong one: e.g., for $J/|J_0|=0.5$ ($T_f=0.167|J_0|$),  $f_p \approx 30$, and for $J/|J_0|=4$, $f_p \approx 3$. It means that the GF remains extremely important for weak disorders, but its influence is 
quite reduced for stronger disorders.

Other result that indicates the presence of two disordered regimes is the  correlation function $\langle \sigma_i \sigma_j\rangle_{intra}$ (Fig. \ref{corr}-b). This intracluster correlation function considers only interactions between spins of the same cluster. It gives information on the spin-spin correlation inside the clusters, but not among different clusters or on the whole system. Figure \ref{corr}-b shows that this correlation is negative for weak disorders. 
In particular, $\langle \sigma_i \sigma_j \rangle_{intra}$ is very close to that one for the pure AF kagome  (see inset of Fig. \ref{corr}-b). However, it becomes positive at low temperatures for stronger disorders. This change in the correlation function suggests that pairs coupled ferromagnetically inside the cluster are energetically favored by disorder. As a consequence, the cluster magnetic moment increases, which can favor the intercluster disordered couplings. These results indicate that a high intercluster disorder overcomes the 
intracluster interactions, supressing the effects of GF.

\section{Remarks and Conclusion}\label{conc}

We study the interplay between GF and disorder by using the Ising kagome lattice in a cluster method. The model adopts a van Hemmen type of disordered interaction between clusters combined with the short-range CCMF approach. This treatment allows us to access thermodynamic information of the disordered kagome lattice and draw phase diagrams to analyze the competition between CSG and conventional orders.

The CCMF method improves the mean-field description of the pure kagome lattice. In particular, the critical temperature for the FE kagome lattice and  the absence of long-range order in the AF case are in good agreement with the exact results \cite{Syozi1951,Kano1953}. The thermodynamic description of the entropy, magnetic susceptibility and spin-spin correlation 
function indicates that the CCMF theory is a valuable tool to study geometrically frustrated systems. 

The presence of disorder in the FE kagome lattice introduces competition between ferromagnetism and CSG state. The FE order is found at low intensities of disorder. However, the CSG phase appears at sufficiently strong disorder. The disorder actives the presence of clusters, in which the intracluster FE interactions introduce a high cluster magnetic moment. This situation enhances the effect of the intercluster disorder that favors the CSG behavior. It means that a critical value of disorder is needed to establish the CSG phase.  

On the contrary, an infinitesimal disorder is able to stabilize the CSG behavior in the geometrically frustrated kagome lattice. In this case the lattice is divided into clusters for any disorder intensity. The entropy goes to zero when the temperature decreases and there is no residual entropy. 
It suggests that any amount of disorder is enough to stabilize the CSG phase instead of the spin-liquid  ground state.
Nevertheless, the frustration parameter is extremely high for weak disorders. These results indicate that the GF of the whole lattice is replaced by geometrically frustrated clusters, which helps to introduce the CSG ground state for an infinitesimal disorder. 

This interpretation is also supported by the freezing temperature as a function of disorder strenght that shows two different behaviors. For weak disorders, the $T_f(J)$ inclination is smaller than in a strong disorder regime. The intracluster spin correlation function indicates that the AF interactions remain important inside the cluster for weak disorders. These frustrated intracluster interactions favor an uncompensated cluster magnetic moment, which is small but sufficient to stabilize the CSG behavior at weak intensities of $J$. As disorder increases, the intracluster AF interactions lose importance and the  spin-spin correlation becomes positive. Thus, a higher cluster magnetic moment is energetically favored by the intercluster disordered interactions. It increases the $T_f(J)$ inclination  at the same time that the frustration parameter  becomes very small. In other words, an infinitesimal disorder is enough to bring a cluster CSG phase in which the GF plays an important role. However, as long as 
disorder increases, the effect of GF is ruled out. Our results suggest that the formation of very small spin clusters can contribute to stabilize a spin glass behavior even at an extremely low level of intercluster disorder. 
 In this context, the GF helps to introduce uncompensated spin clusters, which become hypersensitive to disorder.  Although we study a particular model, the results for weak disorders  can shed some light in the unconventional SG-like behavior in the compound Co$_3$Mg(OH)$_6$Cl$_2$ \cite{Kindo2012}.  
  In particular, the results reported in figures 4 and 5 (a) for entropy and magnetic susceptibility are qualitatively similar to those found in this compound.

\section*{Acknowledgments}

This work was partially supported by the Brazilian
agencies CNPq, FAPERGS, and CAPES.
We thank  L. F. Barquin for helpful discussions.

\bibliography{References}{}
\bibliographystyle{elsarticle-num}
\end{document}